\documentclass[aps,prl,reprint,groupedaddress,superscriptaddress,showpacs]{revtex4-1}
\usepackage{graphicx}
\usepackage{amssymb}
\usepackage{hyperref}

\begin{document}

\title{Induced Superconductivity in Graphene Grown on Rhenium}

\author{C. Tonnoir}
\affiliation{SPSMS, UMR-E 9001, CEA-INAC/UJF-Grenoble\,1, 17 rue des martyrs, 38054 Grenoble cedex\,9, France}
\author{A. Kimouche}
\affiliation{Universit\'e Grenoble Alpes, Inst NEEL, 25 rue des Martyrs, F-38042 Grenoble cedex 9, France}
\affiliation{CNRS, Inst NEEL, 25 rue des Martyrs, F-38042 Grenoble cedex 9, France}
\author{J. Coraux}
\affiliation{Universit\'e Grenoble Alpes, Inst NEEL, 25 rue des Martyrs, F-38042 Grenoble cedex 9, France}
\affiliation{CNRS, Inst NEEL, 25 rue des Martyrs, F-38042 Grenoble cedex 9, France}
\author{L. Magaud}
\affiliation{Universit\'e Grenoble Alpes, Inst NEEL, 25 rue des Martyrs, F-38042 Grenoble cedex 9, France}
\affiliation{CNRS, Inst NEEL, 25 rue des Martyrs, F-38042 Grenoble cedex 9, France}
\author{B. Delsol}
\affiliation{SIMAP, Grenoble INP, 1130 rue de la Piscine, BP 75, F-38402 Saint-Martin-d'H\`{e}res, France}
\author{B. Gilles}
\affiliation{SIMAP, Grenoble INP, 1130 rue de la Piscine, BP 75, F-38402 Saint-Martin-d'H\`{e}res, France}
\author{C. Chapelier}
\email{claude.chapelier@cea.fr}
\affiliation{SPSMS, UMR-E 9001, CEA-INAC/UJF-Grenoble\,1, 17 rue des martyrs, 38054 Grenoble cedex\,9, France}

\date{\today}

\begin{abstract}
We report a new way to strongly couple graphene to a superconductor. The graphene monolayer has been grown directly on top of a superconducting Re(0001) thin film and characterized by scanning tunneling microscopy and spectroscopy. We observed a moir\'{e} pattern due to the mismatch between Re and graphene lattice parameters, that we have simulated with \textit{ab initio} calculations. The density of states around the Fermi energy appears to be position dependent on this moir\'{e} pattern. Tunneling spectroscopy performed at 50 mK shows that the superconducting behavior of graphene on Re is well described by the Bardeen-Cooper-Schrieffer theory and stands for a very good interface between the graphene and its metallic substrate.
\end{abstract}

\maketitle

	Since its discovery in 2004, graphene has attracted a lot of attention because of its unique electronic properties \cite{Geim2007}. In this atomically thin sheet of carbon atoms arranged in a honeycomb lattice, electrons and holes obey to a linear dispersion law and can be described as massless Dirac fermions with a Fermi level coinciding with the Dirac point. This unique situation in condensed matter physics is signaled by an anomalous quantum Hall effect \cite{Novoselov2005}. The relativistic quantum description of graphene has also important consequences in the physics of superconductivity \cite{Beenakker2008}. Although bare graphene is not itself superconducting, it is predicted to acquire superconducting properties when doped with alkaline metal adatoms \cite{Uchoa2007,Profeta2012}. In this situation the Cooper pairing mechanism is either unconventional, with graphene electrons coupling to the metal plasmons, or conventional with electron-phonon coupling between the charge carriers and the different vibrational modes of the carbon and the metal arrays.\\
	Another way to induce superconductivity in graphene is to connect it to a superconductor. In this case, the Cooper pairs are only created in the superconducting material. This proximity effect is governed by the Andreev process which converts electrons and holes of a normal metal into each other, allowing the injection of a Cooper pair into the superconductor. A new situation appears in undoped graphene where the hole of the Andreev process is not retroreflected anymore but undergoes a specular Andreev reflection \cite{Titov2007, Beenakker2008}. Despite an increasing theoretical activity focusing on the superconductivity in graphene and more generally in Dirac electronic systems \cite{Qi2011}, on the experimental side, whereas a tunable Josephson supercurrent has been observed in graphene soon after its discovery \cite{Heersche2007}, the superconducting proximity effect in graphene remains very challenging \cite{Du2008, Ojeda-Aristizabal2009, Kessler2010}.  One reason is the difficulty to prepare a highly transparent interface between graphene and a superconducting metal \cite{Ojeda-Aristizabal2009}. Indeed, while the classical transport experiments are dependent on the transparency $T$ of the interface, the proximity effect ruled by the Andreev reflection is $T^{2}$ dependent.
In this work we demonstrate a new efficient way to induce superconductivity into graphene by directly growing it onto rhenium, taking advantage of both its catalytic and superconducting properties.\\

Epitaxial Re thin films with 30\,nm thickness were grown using electron beam evaporation in a molecular beam epitaxy setup onto polished single-crystal $\alpha$-Al$_{2}$O$_{3}$(0001) wafers which were previously cleaned by annealing under ultrahigh vacuum for 5\,h at 573\,K \cite{Bauer2003}. The Re deposit was conducted at 773\,K at a rate of 8\,{\AA}/min. Reflection high-energy electron diffraction revealed that the Re films are (0001) terminated with $\left\langle 11\bar{2}0 \right\rangle$ crystallographic directions in sapphire aligned to $\left\langle 1\bar{1}00 \right\rangle$ ones in Re. The Re/sapphire samples were then transferred to a second ultrahigh vacuum system equipped for graphene growth. The Re(0001) surface was annealed at 1073\,K for 30\,min. Then, the Re film was enriched with carbon and cleaned from residual oxygen by exposing its surface to ethylene (5$\times$10$^{-8}$\,mbar) at 1263\,K for 20\,min. During this step, the ethylene molecules are dehydrogenated and resulting C atoms dissolve into the bulk of the sample. The temperature was then slowly decreased to 973\,K in 50\,min, in order to lower the carbon solubility in Re. This results in a progressive depletion of C atoms in the bulk and their segregation on the surface \cite{Yu2008, Kimouche}. A monolayer of graphene fully covering the Re grains is thus formed.
An alternative growing process has been reported recently by Miniussi and coauthors \cite{Miniussi2011} which relies on a sequential chemical vapor deposition process. These authors deduced, from microdiffraction experiments with low-energy electrons, the presence of a moir\'{e} pattern, due to a mismatch of the lattice parameters of graphene ($a_\mathrm{C}$) and Re ($a_\mathrm{Re}$). In this Letter, we report the first real space observation of this superstructure with scanning tunneling microscopy (STM).\\

\begin{figure}[h!]
\includegraphics[width=1\columnwidth]{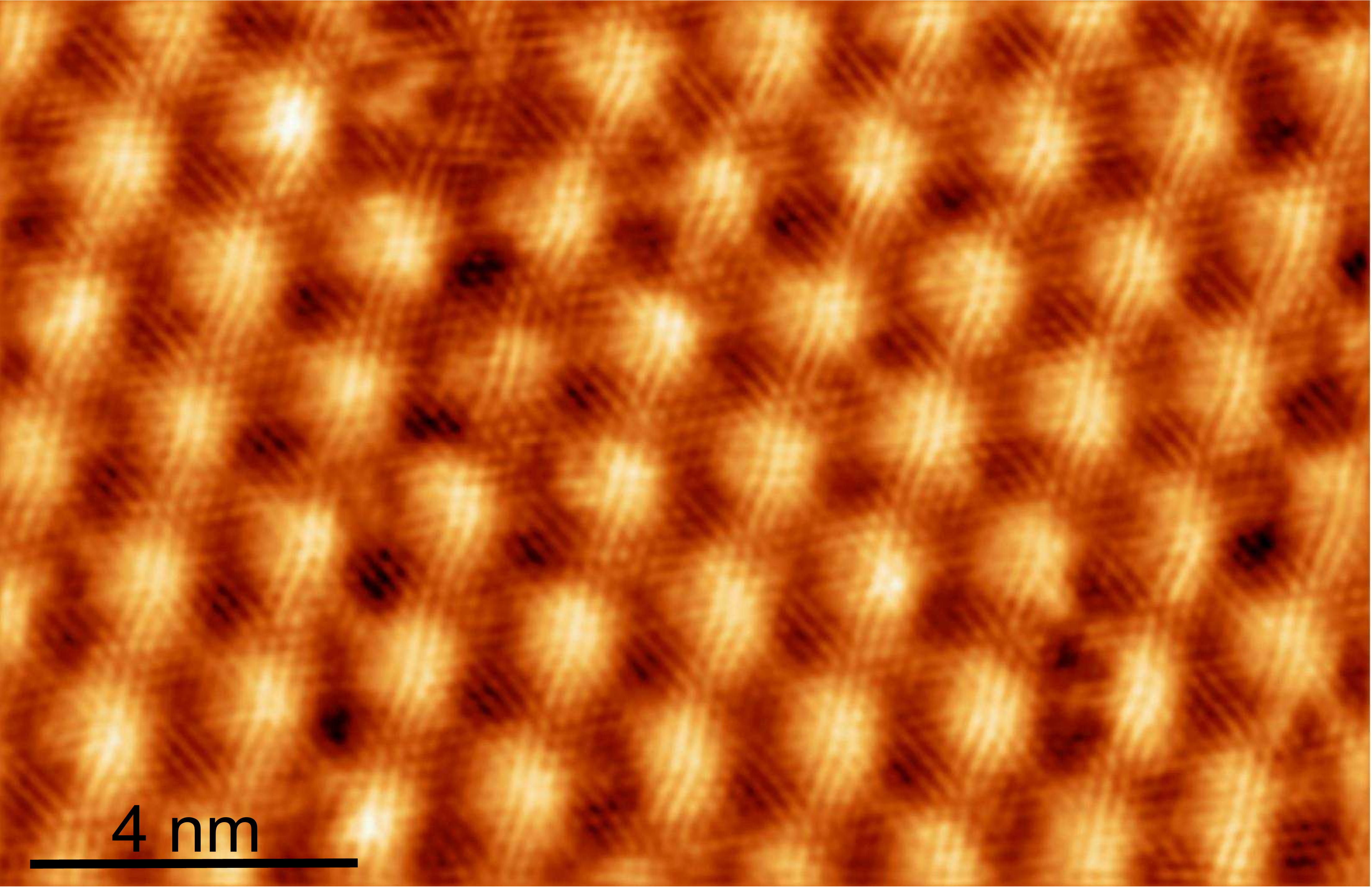}
\includegraphics[width=1\columnwidth]{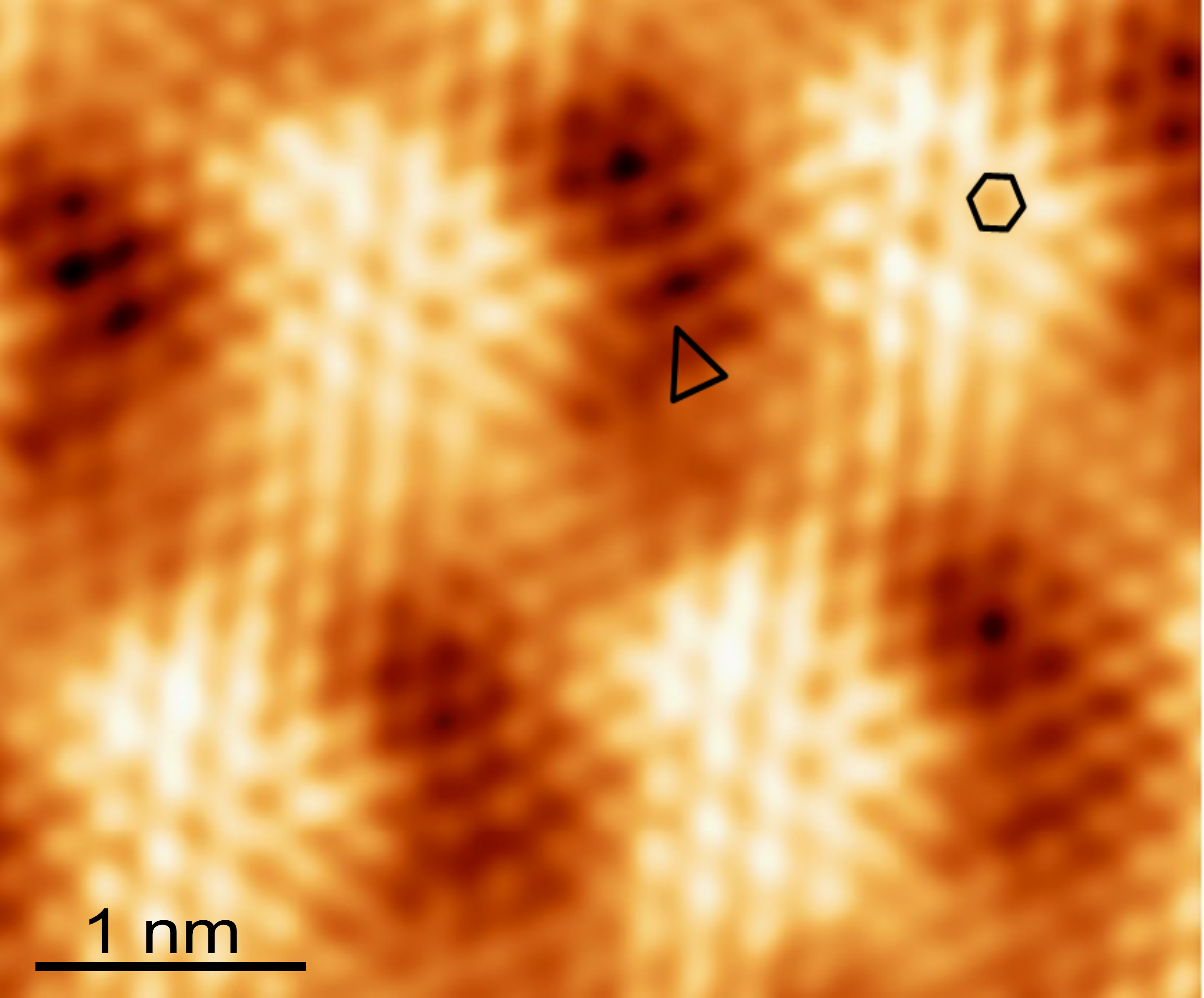}
\caption{Top: STM image showing the moir\'{e} pattern due to the lattice paramater mismatch of graphene and Re, measured with a bias voltage $V_\mathrm{bias}$\,=\,224\,mV and a tunneling current I\,=\,31\,nA. Bottom: STM image exhibiting an atomic periodicity of order six on the hills of the moir\'{e} and three in the valleys, measured for $V_\mathrm{bias}$\,=\,10\,mV and $I$\,=\,10\,nA.}
\label{moire} 
\end{figure}

The top image on Fig.\,\ref{moire} shows this moir\'{e} with a periodicity of 1.9\,$\pm$\,0.1\,nm, which corresponds to a moir\'{e} with carbon zigzag rows aligned to Re dense packed ones, and a (7:8) superstructure, i.e. eight carbon rings matching seven Re atoms. This moir\'{e} is different from the (9:10) cell calculated in Ref. \cite{Miniussi2011}. This discrepancy could originate from the different methods used to grow graphene in the two experiments. Our superstructure goes along with a compressed graphene cell where the graphene lattice parameter is estimated to be 2.42\,\AA, a slightly smaller value than the one in bulk graphite (2.46\,\AA). On the upper image of Fig.\,\ref{moire}, we can see that the atomic rows of graphene are aligned with the moir\'{e} direction. Nevertheless, we also observed spots where the two lattices are rotated. On the example shown in the bottom image of Fig.\,\ref{moire}, the angle between the carbon atomic rows and the moir\'{e} high symmetry directions is $\alpha\sim$\,15$\ensuremath{^\circ}$, which corresponds to a rotational angle between the Re and the graphene lattice of $\beta\simeq\alpha$($a_\mathrm{Re}-a_\mathrm{C})/a_\mathrm{Re}\sim$\,2$\ensuremath{^\circ}$. We measure an apparent moir\'{e} corrugation that is tip dependent but can be as high as 1\,{\AA}. This indicates strong coupling between graphene and Re as compared to other metallic substrates \cite{Batzill2012, Voloshina2012}, similar to what is observed for graphene grown on ruthenium \cite{Marchini2007, VazquezdeParga2008, Pan2009, Gyamfi2011}.
Images with atomic resolution (see Fig.\,\ref{moire} bottom image) show sixfold and threefold periodicity on the hills and valleys of the moir\'{e}, respectively, which suggests that the graphene layer is less strongly coupled to the Re substrate on the hills than on the valleys. In order to corroborate these observations, we performed numerical simulations of the moir\'{e} structure.\\

\begin{figure}[h!]
\includegraphics[width=1\columnwidth]{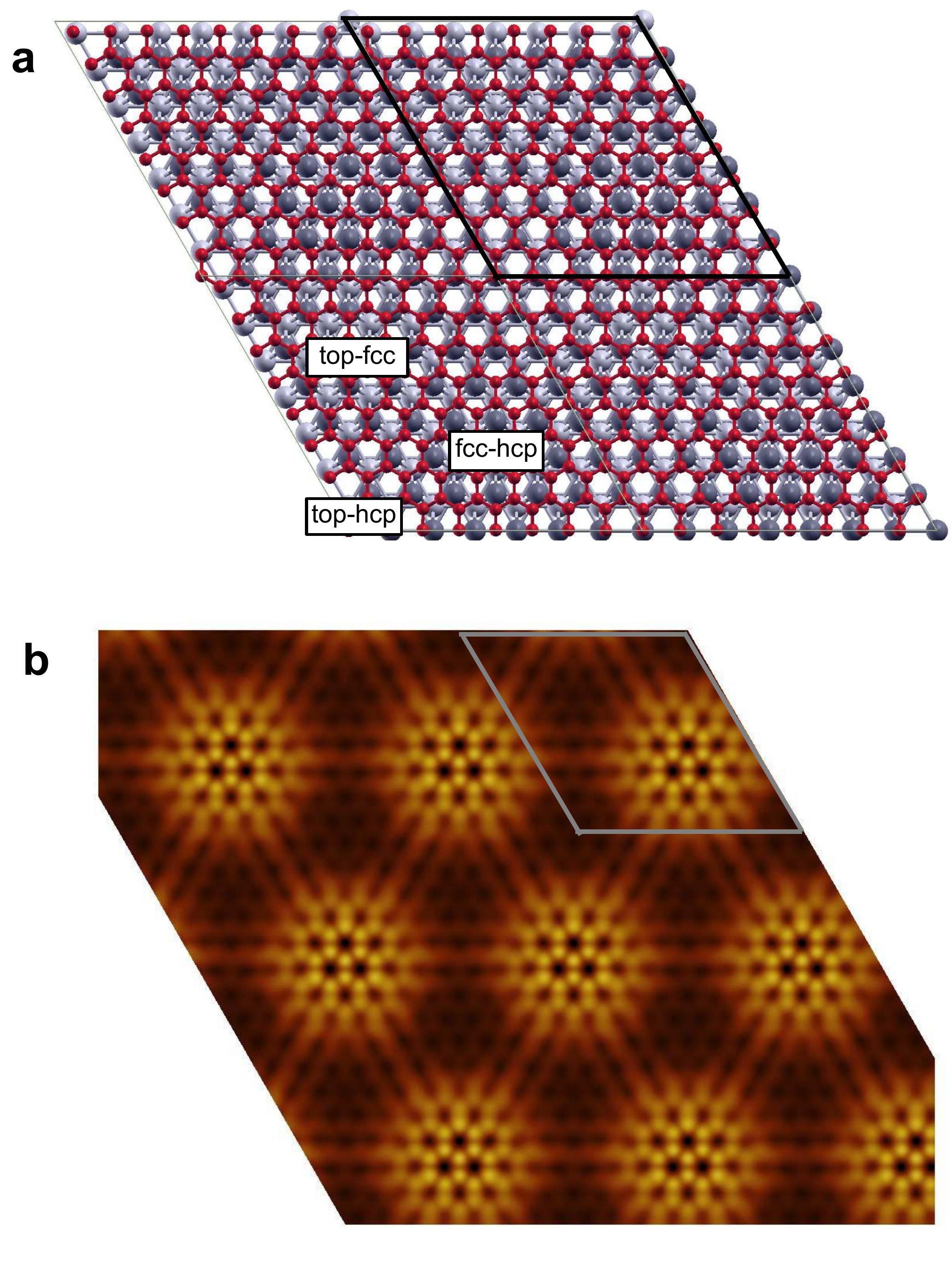}
\caption{(color online).
{\it Ab initio} calculation results for the graphene\,(8$\times$8)\,-\,Re\,(7$\times$7) interface: a) schematic view from top, b) cross section of the square modulus of the wave function integrated between $E_\mathrm{F}$ and $E_\mathrm{F}$\,+\,0.5\,eV.}
\label{Fig_theo} 
\end{figure}

{\it Ab initio} calculations have been carried out with the code VASP \cite{VASP}. The PAW approach \cite{PAW}, PBE functionnal \cite{PBE} and Grimme corrections to van der Waals interactions \cite{Grimme} have been used. The slab used to describe the system contained five Re layers, one graphene layer, and a 10\,\AA -thick vacuum space on top. The Re plane in the middle (third plane) of the slab was fixed while all the other atoms were allowed to relax \cite{sup-mat}.  The lateral size of the supercell has been fixed to the geometry observed experimentally and corresponds to a (7$\times$7) cell for Re and a (8$\times$8) cell for graphene. Calculations were performed with one $k$ point, the supercell $K$ point to get a precise description of the graphene low energy states. After convergence, residual forces were lower than 0.025\,eV/\AA.\\
Our calculations show that the graphene layer is buckled with regions where the C atoms are close to Re ones and regions where they lay much higher (Fig.\,\ref{Fig_theo}). The first regions correspond to strong graphene-Re interaction (top hcp and top fcc - see definition in \cite{Batzill2012}) with the formation of covalent bonds and an electron transfer from Re to C atoms \cite{sup-mat}. The second regions correspond to a much weaker interaction (hcp-fcc). This is consistent with our experimental observations and with the description given by Miniussi $et al.$ \cite{Miniussi2011} but for a (9:10) cell. Figure\,\ref{Fig_theo} (b) shows the square modulus of the wave function integrated between Fermi level ($E_\mathrm{F}$) and $E_\mathrm{F}$\,+0.5\,eV. This cross section is taken just above the highest graphene atom. It shows bright protuberances that correspond to hcp-fcc stacking regions and lines. These calculations provide good agreement with our STM images of Fig.\,\ref{moire}.\\

\begin{figure}[h!]
    \includegraphics[width=1\columnwidth]{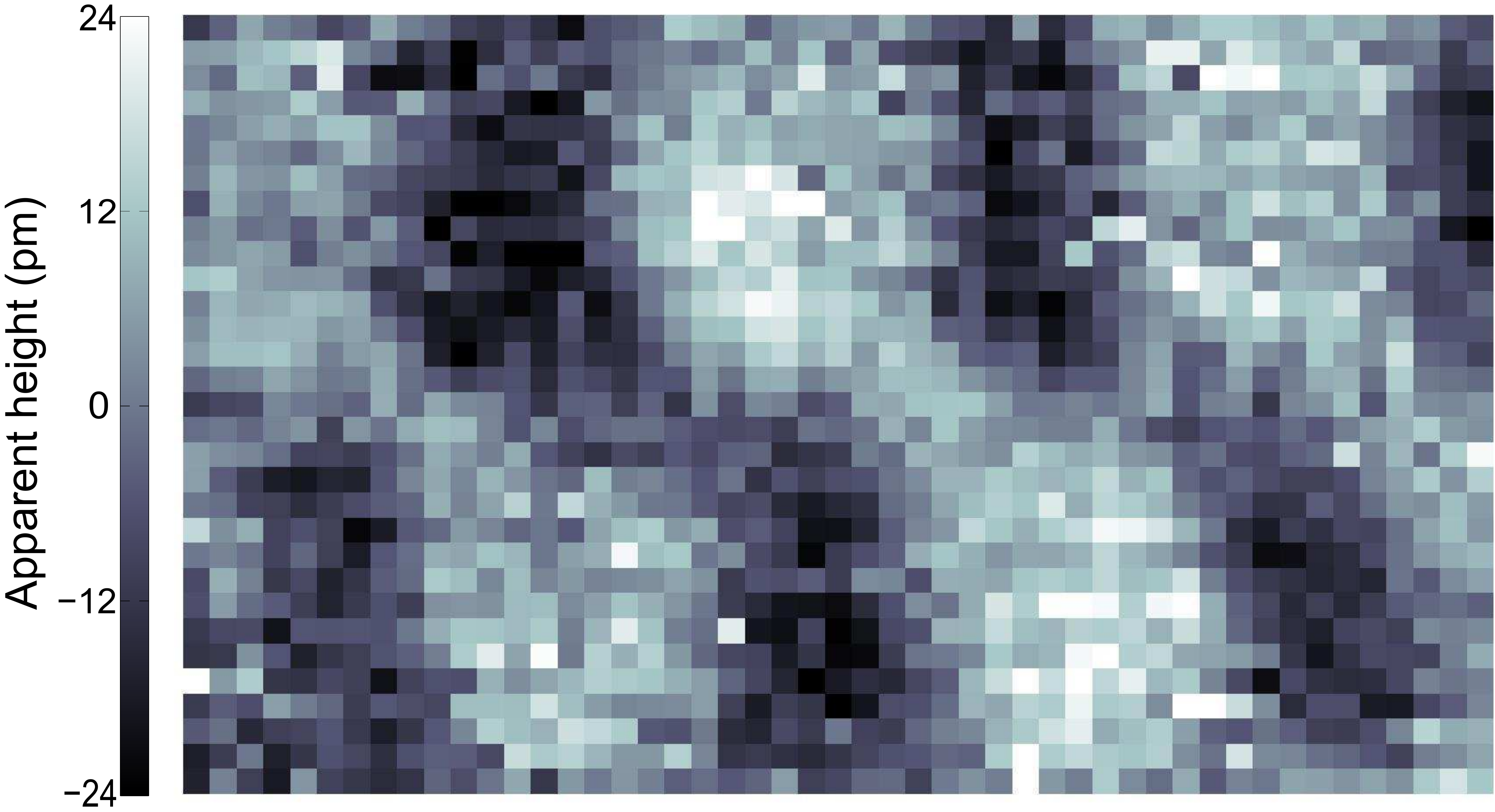}
    \includegraphics[width=1\columnwidth]{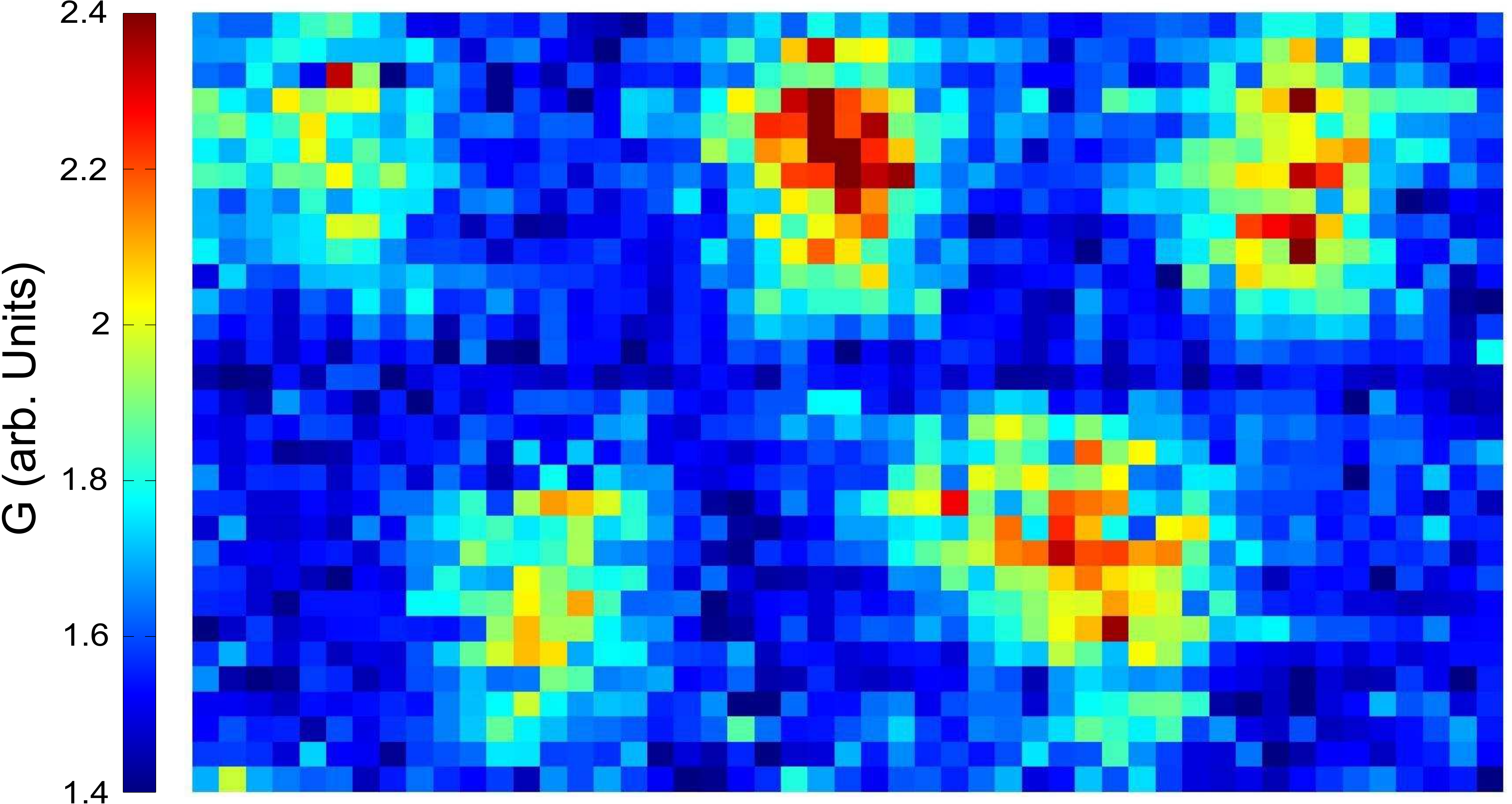}
    \includegraphics[width=1\columnwidth]{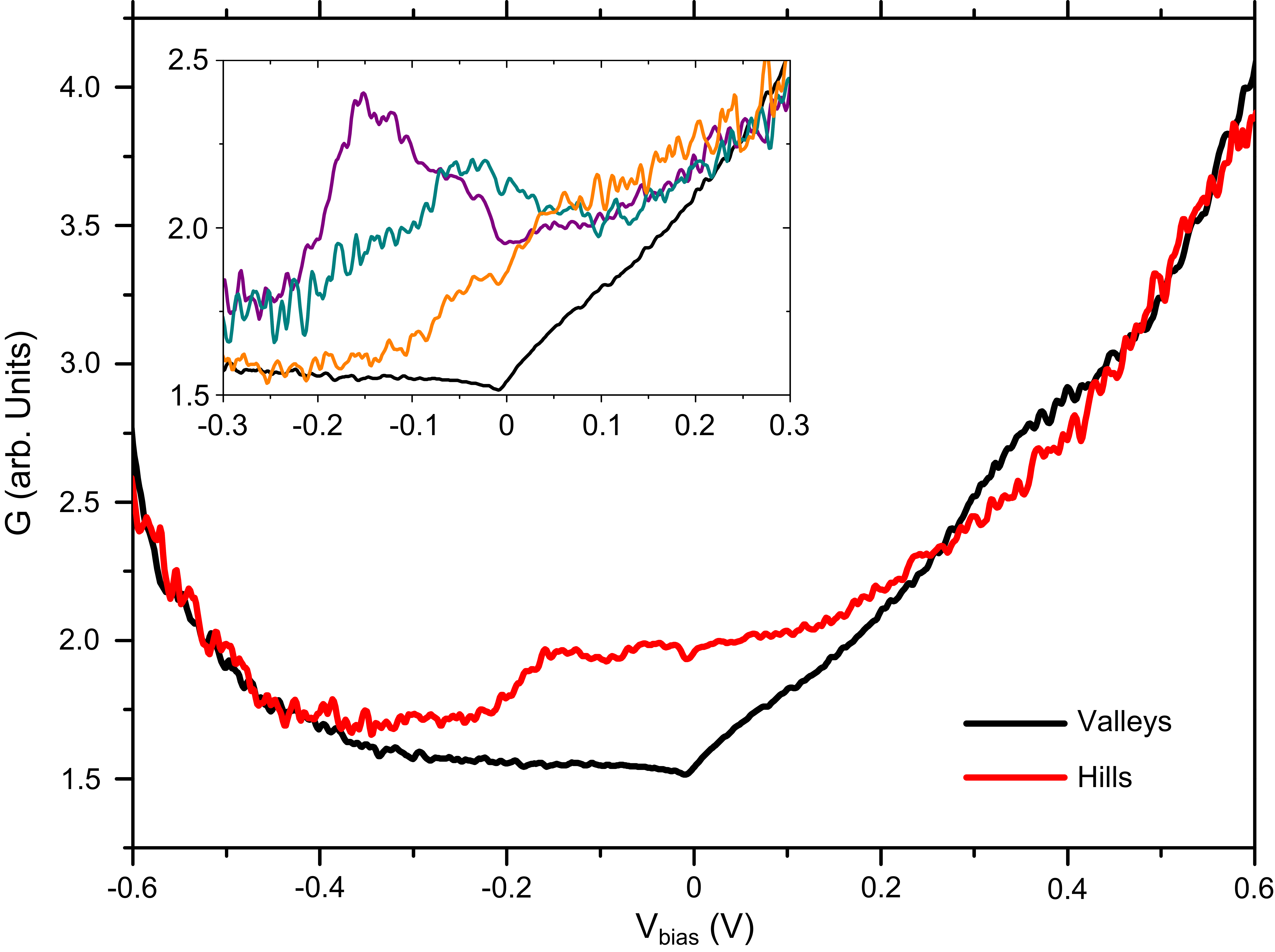}
	\caption{(color on line). Top: Topographic image (5$\times$3\,nm$^{2}$) measured at $V_\mathrm{bias}$\,=\,800\,mV and I\,=\,5\,nA. Middle: Conductance map at the Fermi energy. Bottom: Spectra averaged over the hills (red) and valleys (black). Inset: zoom in energy of spectra averaged on individual hills (colors) and the spectrum averaged on the valleys (black).
    \label{CITS}}
\end{figure}

Another signature of the coupling strength between graphene and the Re substrate can be found in the local density of states (DOS). This is highlighted in Fig.\,\ref{CITS} where we show a topographic image of the moir\'{e} and a simultaneously recorded map of the differential conductance $G=dI/dV$ measured at $E_\mathrm{F}$ [i.e. at zero tip-sample bias ($V_\mathrm{bias}$)]. These data have been acquired in a dilution refrigerator where the STM was cooled down to 50\,mK. The spectroscopic signal was obtained by a lock-in technique with an rms modulation voltage of 5\,mV. On this spectroscopy map we observe enhanced conductance on the hills. This contrast can be understood by recording a full spectrum on hills and valleys (see Fig.\,\ref{CITS} bottom graph). The tunneling spectra show a typical semimetal behavior, similar to what was obtained for graphene/Ru(0001) \cite{Pan2009}, in contrast to the V-shaped DOS of a decoupled graphene layer \cite{Andrei2012}. This indicates an overall strong coupling between graphene and the Re substrate. However, the averaged tunneling conductance measured on the hills is larger than the one measured in the valleys between -400\,mV and +200\,mV. These strong electronic effects are likely to play a central role in the observed moir\'{e}. Surprisingly, in this energy window, spectra acquired on different hills display a maximum for different energies (see Fig.\,\ref{CITS} inset in bottom graph). These spectroscopic variations between different hills could come from a slight misalignment between the crystallographic directions of the graphene and the Re. The resulting incommensurability in the moir\'{e} would then lead to nonequivalent hills regarding their electronic properties. This scenario needs further investigation to be verified.\\

\begin{figure}[h!]
\includegraphics[width=1\columnwidth]{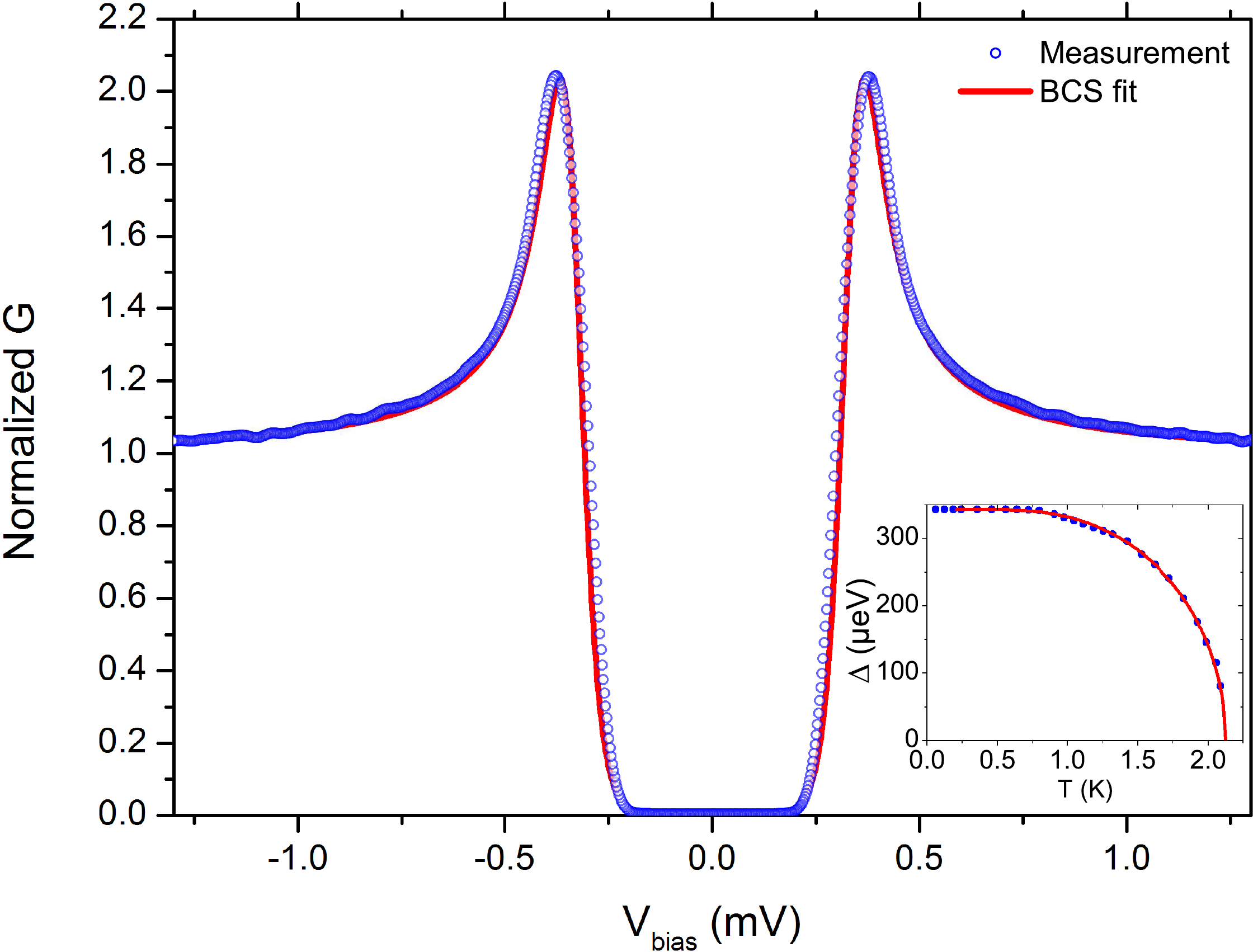}
\caption{Normalized differential conductance G measured at 50\,mK. Inset: temperature evolution of the superconducting gap.}
\label{gap} 
\end{figure}

Tunneling spectroscopy performed at 50\,mK and low bias voltage with a lock-in rms modulation voltage of 40\,$\mu$V revealed a very uniform superconducting gap $\Delta$\,=\,330\,$\pm$10\,$\mu$eV, which can be well fitted with the standard Bardeen-Cooper-Schrieffer (BCS) \cite{Bardeen1957} (Fig.\,\ref{gap}). The thermal dependence also follows very accurately the BCS model (see inset in Fig.\,\ref{gap}). We can deduce a superconducting transition temperature $T_\mathrm{C}\simeq$\,2.1\,K and a ratio $\Delta$/$T_\mathrm{C}$\,=\,1.88. This type of measurement is a very sensitive probe of the quality of the interface between the normal "metal" ($N$), here graphene, and the superconductor ($S$). Indeed, the size of the gap induced by proximity effect in $N$ depends on the barrier height between $N$ and $S$. In absence of this barrier, the superconducting gap in $S$ and the induced one in $N$ are equal, the latter being reduced for a less transparent interface \cite{McMillan1968}. As a matter of fact, we measured $\Delta$\,=\,255\,$\mu$eV and $T_\mathrm{C}$\,=\,1.6\,K on bare Re films with a similar $\Delta$/$T_\mathrm{C}$ \cite{Dubouchet}. The higher values of $\Delta$ and $T_\mathrm{C}$ measured on the graphene layer can be explained by a DOS modification of Re when carbon atoms are dissolved into its bulk, together with a perfectly transparent interface between graphene and Re. Whereas we have shown in the first part of the paper that the coupling strength between graphene and rhenium varies at the moir\'{e} scale, we did not observe any significant spatial variation of the superconducting spectrum. This is not surprising since the period of the moir\'{e} is much shorter than the superconducting coherence length of 24\,nm measured in Re.\\

\begin{figure}[h!]
    \includegraphics[width=1 \columnwidth]{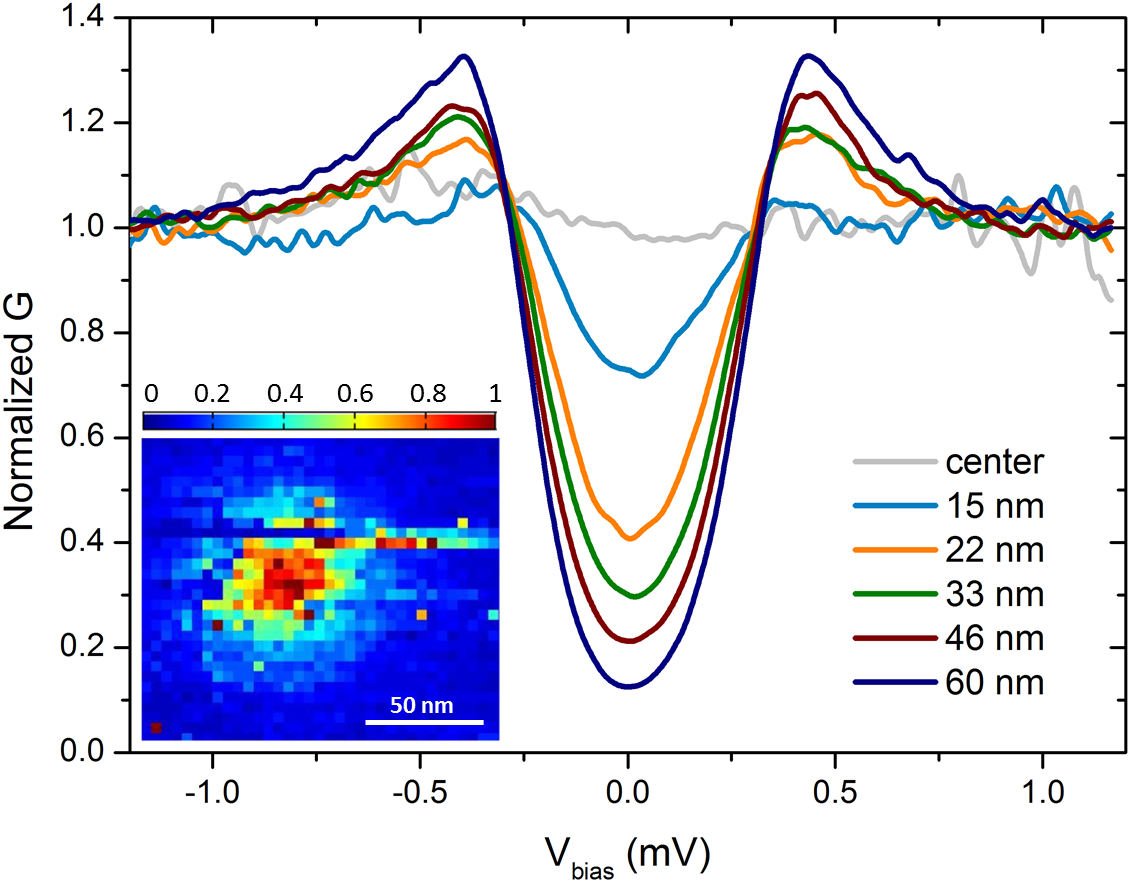} 
    \caption{Closing of the superconducting gap in the vortex core region. Inset: Conductance map of the vortex core (155$\times$135\,nm$^{2}$) at the Fermi energy.
    \label{vortex}}
\end{figure}

A second method to estimate the barrier strength between a type II superconductor $S$ and $N$ consists in measuring the magnetic vortex core radius in $N$, which is expected to increase for decreasing interface transparencies \cite{Golubov1996}. The conductance map at $E_\mathrm{F}$ acquired under a perpendicular magnetic field of 57.5\,mT together with the spatial evolution of the spectra as a function of the distance to the vortex center are displayed in Fig.\,\ref{vortex}. The observed vortex core is slightly smaller than what we had previously observed in bare Re films \cite{Dubouchet}, which confirms the good transparency between graphene and Re. Interestingly, two lines of the STS image display the vortex core shifted from the vortex center at rest. This is a well-known effect due to the metastability of the vortex between two pinning centers \cite{Dubois2008, Hoogenboom2000}.\\

	In conclusion, we have observed the moir\'{e} pattern of epitaxially grown graphene on Re, whose superstructure consists in (8$\times$8) carbon atoms over (7$\times$7) Re atoms. The spectroscopic measurements and \textit{ab initio} calculations demonstrate a strong coupling between graphene and Re. This system therefore constitutes a building block to design hybrid superconducting nanostructures where the good transparency of the $SN$ interface is of primary importance. In these hybrid devices the $N$ areas can be obtained by decoupling locally the graphene from the substrate in a controlled way. Among other methods, this could be achieved by intercalation \cite{Sutter2010, Lizzit2012,Granas2012}.\\

\begin{acknowledgments}
	The authors acknowledge the Grant No.\,ANR-BLANC-SIMI10-LS-100617-12-01 - Supergraph - from the "Agence Nationale de la Recherche", the Grenoble Nanosciences Fondation grant SuperNanoCharac, DISPOGRAPH, and the European Project FP7-NMP-2009-SMALL-3 Grenada. Theoretical work was performed using HPC resources from GENCI-IDRIS (Grant No.\,2013-097015). We thank Andrea Locatelli, Tevfik Onur Mentes and Nicolas Rougemaille for fruitful discussions.
\end{acknowledgments}



\end{document}